# SciNote: Collaborative Problem Solving and Argumentation Tool

Janet Rafner, Arthur Hjorth, Carrie Weidner, Shaeema Zaman Ahmed, Christian Poulsen, Clemens Klokmose, Jacob Sherson
janetrafner@mgmt.au.dk, arthurhjorth@phys.au.dk, cweidner@phys.au.dk,
shaeema@phys.au.dk, cbp@mgmt.au.dk, clemens@cavi.au.dk, sherson@mgmt.au.dk
Aarhus University, Aarhus, Denmark

**Abstract:** As educators push for students to learn science by *doing* science, there is a need for computational scaffolding to assist students' evaluation of scientific evidence and argument building. In this paper, we present a pilot study of SciNote, a CSCL tool allowing educators to integrate third-party software into a flexible and collaborative workspace. We explore how SciNote enables teams to build data-driven arguments during inquiry-based learning activities.

## Introduction and theoretical considerations

The last three decades have seen an ever-increasing call for educational activities in which learners *do* science instead of *learning about* science. A growing number of initiatives support the general public's participation in research, but there is an urgent need for tools to facilitate these processes in both formal and informal learning contexts. When moving *beyond mere data-gathering, scientific argumentation* is key to actively engaging students in epistemological and methodological considerations. The CSCL community has sought to support this inquiry-oriented turn by providing computational scaffolding (Quintana et al, 2004) to learners engaged in argumentation in a variety of ways: AL (Wambsganss et al, 2020) gives learners real-time feedback on written argumentation; Critical Thinking (Sun, Yuan, Rosson, Wu, & Carroll, 2017) supports dyads discussing a particular proposition; and ThoughtSwap (Dickey-Kurdziolek, Schaefer, Tatar, & Renga, 2010) assists students in discussing the fitness of an answer to a question. In this pilot study, we introduce SciNote, a CSCL tool that both supports collaborative problem solving by integration of scientific data generated from third-party tools and assists learners in evaluating scientific evidence and building scientific arguments.

## Introducing SciNote and pilot study

SciNote is an online tool, built on Webstrates (Klokmose, Eagan, Baader, Mackay, & Beaudouin-Lafon, 2015**)**. The goal of SciNote is to scaffold students' inquiry-based learning experience through two modes: *mode 1:* problem solving, sharing and building upon other's solutions to collectively explore the solution space and arrive at the best possible solutions (Fig. 1a, above dashed line) and *mode 2:* collective argumentation in which insights from mode 1 can be refined and aggregated into a deep phenomenological investigation of the underlying topical challenge (Fig. 1a, below dashed line). To do so, SciNote contains three specific design features: 1) data integration with third party software, 2) 'mini-papers' that consist of data (numerical results or arguments) and 3) 'citations', a means for students to refer to each other's mini-papers as they collectively build arguments. In Figure 1a we show a screenshot of SciNote. All of a team's mini-papers are visualized as small rectangles showing the title of the evidence and a link to all mini-papers that it cites (or that cite it). Clicking a mini-paper opens it so students can see its data and argument. In this pilot study, we focus on investigating whether such a tool can successfully form the backbone of an inquiry-based learning activity (see Fig. 1b, inset).

## Methods and Context

The study was done with 17 students in a graduate-level quantum mechanics course at a Danish university. We integrated the SciNote backend with Quantum Composer (QC), a quantum physics education and research tool (Ahmed et al, 2020). Students used QC to explore a difficult physics problem with an enormous number of possible solutions; any solution can be evaluated in terms of its proximity to an ideal solution by a score ranging from 0 (a poor solution) to 1 (a perfect solution). Onboarding took place via a 30-minute in-person presentation introducing students to SciNote and the challenge. Over the next hour, the students worked on the challenge in four teams. Students could talk to each other during the activity, but sharing of the details of a given solution (encoded as a JSON file) was only possible using the real-time-updating SciNote interface. To encourage reflection on the process of scientific argumentation, each team was explicitly evaluated on the complexity of their final joint argumentation and the extent to which they had used citations to acknowledge and build on each other's contributions.

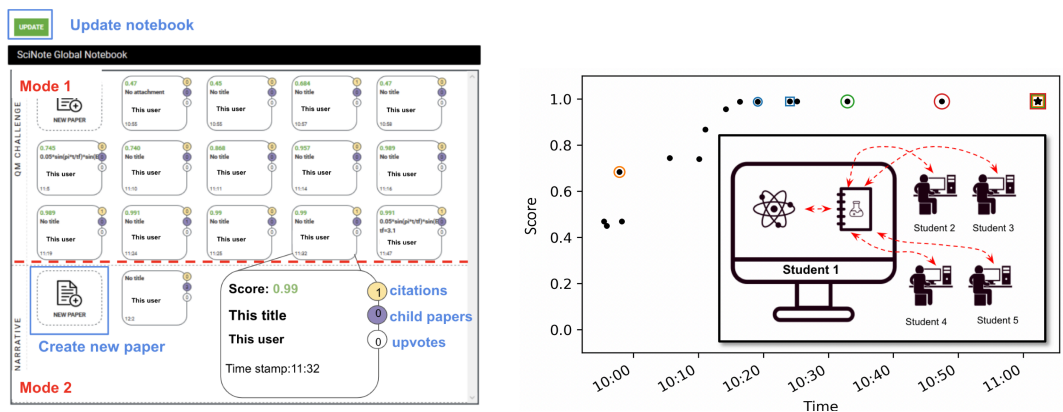

Figure 1. (a, left) *SciNote interface*: screenshot from Team 1 during the study, including an inset to explain each feature (b, right) *Iteratively improving on each other's results:* Team 1 score vs. time, with rings (boxes) around cited (citing) papers, culminating in a final analysis paper (star). (inset) Collaborative quantum problem solving using the SciNote interface.

All teams' scores increased roughly monotonically over time, indicating that teams built upon their previous solutions via citations and intra-team oral communication. For example, Team 1 submitted four papers at the beginning of the exercise (Fig. 1b), and all further work built on the best of these (score = 0.72), as the score never dropped below this level for the rest of the activity; other teams' results were similar. Finally, teams with more citation activity achieved higher scores and demonstrated higher levels of physical understanding in their analyses. The number of citations (final score) for each team were (4, 0.991), (1, 0.72), (9, 0.88), (1, 0.785). Therefore, while this is a pilot study with a small number of participants, these are indications that increased engagement with SciNote led to better knowledge exchange and ultimately better results.

## Conclusion and outlook

Here, we demonstrate that SciNote can facilitate collective, complex problem-solving activities with third party tools, previously only possible with tedious manual file sharing. Furthermore, we see indications that SciNote can help bridge the gap between such activities and collective scientific argumentation. One student notes:

> It seems to work great with these 'investigative' types of problems/challenges but I have rarely seen such investigative problems during my normal studies. Perhaps it could be an idea to discuss with other teachers/professors at the university, if they want to make such investigative problems and create some (fun) challenges.

Future work will explore how SciNote can be used in other scientific contexts, including the social sciences, in other educational contexts, in large-scale studies, and in data integration using different third-party interfaces.